\documentclass[aps,prl,groupedaddress, superscriptaddress]{revtex4}  
\usepackage{graphicx}  
\usepackage{dcolumn}   
\usepackage{bm}        
\usepackage{amssymb}   
\usepackage{sidecap}   
\begin{document}

\widetext
\title{Polychromatic polariton laser selector switch}

\affiliation{ICMP, Ecole Polytechnique F\'ed\'erale de Lausanne (EPFL), 1015 Lausanne, Switzerland}

\author{G.~Grosso} \affiliation{ICMP, Ecole Polytechnique F\'ed\'erale de Lausanne (EPFL), 1015 Lausanne, Switzerland}
\author{S.~Trebaol} \affiliation{ICMP, Ecole Polytechnique F\'ed\'erale de Lausanne (EPFL), 1015 Lausanne, Switzerland} 
\author{M.~Wouters} \affiliation{TQC, University of Antwerp, Universiteislpein 1, 2610 Antwerpen, Belgium}
\author{F.~Morier-Genoud} \affiliation{ICMP, Ecole Polytechnique F\'ed\'erale de Lausanne (EPFL), 1015 Lausanne, Switzerland}
\author{M.T.~Portella-Oberli} \affiliation{ICMP, Ecole Polytechnique F\'ed\'erale de Lausanne (EPFL), 1015 Lausanne, Switzerland}
\author{B.~Deveaud} \affiliation{ICMP, Ecole Polytechnique F\'ed\'erale de Lausanne (EPFL), 1015 Lausanne, Switzerland}

\vskip 0.25cm


\maketitle

{\bf 
Integration of optical elements into scalable chips has been at the center of a large effort in recent years \cite{barkai2007,miller2010,beausoleil2008}. Concurrently, the separation between the diverse functions, namely switches, detectors or emitters \cite{reed2010,takahashi2013,volz2012} increases significantly the final number of components on chip. 
Such technical limitations may be overcome by introducing agile devices able, for example, to simultaneously detect, process and emit a coherent signal.  Such a pathway has been explored with different approaches \cite{obrien2009} that bear advantages and drawbacks. Polaritons have often been proposed as promising candidates for multifunctional devices  \cite{menon2010}. Here we present an optical switch based on polariton lasing. An incident monochromatic signal is channeled into several polariton laser beams at different wavelengths by a novel relaxation mechanism which combines bistability, phonon interactions, long polariton lifetime and bosonic stimulation. 
We demonstrate spin logic operations conserving the original polarization state that is fully imprinted onto the coherently emitted signals. }\newline

In the recent past, novel approaches to optical switches have been proposed in the field of excitons in semiconductors. Excitons are electron-hole pairs with the key advantage to easily couple with light. This peculiar feature brings them to the frontline for the realization of all-optical and optoelectronic devices \cite{miller1984,grosso2009}. Other appealing features arise when excitons are strongly coupled with light in semiconductor microcavities. The resulting particles are known as exciton-polaritons. Polaritons have a very light mass, due to the photonic content, and interact among themselves due to their excitonic part giving gives rise to spectacular nonlinear effects such as solitons \cite{grosso2012} and optical bistability \cite{paraiso2010}.  On top of that, the polariton density and spin can be easily manipulated by optical means. The combination of all those unique properties favored the realization of a large variety of devices such as transistors \cite{anton2012,ballarini2013}, ultrafast memories \cite{cerna2013}, spin switches \cite{amo2010} and led to the observation of Bose-Einstein condensation \cite{kasprzak2006} and polariton lasing \cite{deng2003,schneider2013} up to room temperature \cite{christopoulos2007}. The accumulation of particles at the lower energy state, necessary to achieve the phase transition, has been achieved so far by means of either non-resonant excitation or parametric scattering amplification. Employing these techniques, the resulting coherent state is governed by the intrinsic nature of the system and the control on the final coherent emissions is still an open challenge.\newline

In this letter we report on a new relaxation mechanism for polaritons and its application in an innovative device allowing for all-optical logic operations. We take advantage of the one-to-one connection between intensity (polarization) of the incident beam with the density (spin) of the trapped polariton gas to drive multivalued output states which are wavelength and polarization sensitive. A truth table with nine different output states is obtained.
Polariton lasing is demonstrated by linewidth narrowing of the polaritons states \cite{bajoni2008} and the general experimental findings are validated in the framework of the generalized Gross-Pitaevskii equation.\newline

The device is engineered in a GaAs microcavity (Fig.1a) in which polariton lasing at different colors is realized by confining polaritons inside photonic potential traps (called mesa) giving rise to discretized energy states \cite{kaitouni2006}.\newline 
Fig.1b shows the calculated photonic modes inside a 3 $\mu$m quasi-circular mesa \cite{lugan2006}. Five confined states (CS0-4) appear inside the trap (blue lines) at different energies and with different momentum distribution. At energies greater than the confinement depth, the photonic dispersion shows the typical parabolic shape of planar microcavities. Fig.1c is the experimental polariton dispersion of the system in the linear regime. The slightly elliptical shape of the mesa breaks the rotational symmetry of the system and lifts the degeneracy of the confined states CS1 and CS2 creating two new eigenvalues for each of them \cite{nardin2010}. It is worth to note that all the photonic modes of the mesa are coupled with the exciton ($X$) at different detuning $\delta_{CS,j} = E_{CS,j} - E_X$. Therefore the resulting confined polariton modes (labeled in the following with S$j$ as shown in Fig.1c) have different lifetime due to the diverse photonic and excitonic contents.
The latter is found to be 0.87\%, 0.28\% and 0.14\% for S3, S2-S1 and S0 respectively. This results in a lifetime for S3 which is an order of magnitude higher than S0.\newline

The relaxation mechanism inside the mesa is studied using photoluminescence excitation (PLE) in the linear regime (P = 100 $\mu$W). Fig.1d shows the PL spectra (right axis) as a function of the excitation energy (bottom axis). 
The PL emission due to resonant excitation is stronger for the lower confined states (S0-2) and almost no emission is observed for S3 as expected from its high excitonic content. This feature governs also the relaxation mechanism. When exciting S1 and S2 no relaxation toward S0 occurs. On the contrary, a strong increase appears when exciting S3. This behavior is related to the different polariton lifetime. Despite of the large energy separation between the states, the phonon assisted relaxation can occur due to the relatively long lifetime of S3 and to the enhanced overlap between the phonon and the confined polariton wavefunction \cite{paraiso2009}.\newline

The main idea of this work is to exploit the efficient relaxation from S3 and imprint its strong nonlinear behavior onto the lower states. Indeed, S3 displays a nonlinear optical bistability due to polariton-polariton
interactions \cite{paraiso2010,gippius2007}. Polaritons are selectively injected in the center of the mesa with a momentum $k_{\parallel}=0$. Therefore the relaxation is predominant toward confined states with a wavefunction peaked around $k_{\parallel}=0$  \cite{bajoni2008}.
In order to create a bistable behavior with nonlinear optical jump, the pump laser is slightly detuned with respect to state S3 ($\delta$=0.3 meV). An energy resolved intensity map as a function of the pump power is plotted in Fig.2a in saturated color scale. The laser beam is prepared in circular polarization ($\sigma^+$) and the emission is recorded for co-circular polarization. The normalized intensity for the polariton states is then extracted and displayed in Fig.2b. At low power, the S3 intensity follows the expected linear behavior. The increasing density leads to a blueshift of the polariton states because of the polariton-polariton interaction. Around P=8 mW, S3 gets in resonance with the laser showing an abrupt jump in the intensity due to the more efficient coupling with the laser. The nonlinear jump is then imprinted from S3 to S2, S1 and S0. The different behavior at the threshold of the lower states depends on the energy separation with respect to S3 and the momentum dispersion, favoring S2. Moreover, the macroscopic occupation of the states is responsible for bosonic stimulation which is activated at 8 mW when the density threshold is reached in all the lower states. Around P=20 mW, another nonlinear increase is observed for S0 accompanied by a decrease of the S1 and S2 occupancy. This unusual behavior is the result of the interplay of bosonic stimulated relaxations among the different states. 
The linewidth of all the emissions narrow at the first threshold (Fig.2c, 4SM and 6SM) approaching the size of S3 which is limited by the resolution of the spectrometer. This observation shows the onset of phase coherence in the confined states which all behave as polariton lasers. This behavior is enriched by the conservation of the spin during the relaxation mechanism.\newline
Fig.2d shows the same power map of Fig.2a but now the $\sigma^+$ emission is filtered out although it is still visible

due to its strong intensity compared to $\sigma^{-}$. This helps us to visualize how  $\sigma^+$ blueshifts contrarily to $\sigma^-$. Moreover the linewidth of $\sigma^-$ does not experience any narrowing after the threshold but increases instead due to homogeneous broadening (Supplementary Material).\newline
The experimental results and interpretations are supported by theory. It is in fact possible to model our system in the framework of the Gross-Pitvaeskii equation, widely used to describe polariton systems \cite{carusotto2013}. The confined states of the mesa can be treated as separate polariton states coupled through the relaxation process. The coupled equations for the field amplitude of S3 ($\psi_3$), the density of S2 ($n_2$), S1 ($n_1$) and of S0 ($n_0$) read
\begin{center}
\begin{eqnarray}
i \hbar \frac{\partial}{\partial t} \psi_3 = \left[ g\sum_{j=0}^{3} n_j - \frac{i}{2} \left(  \gamma_3 + \sum_{l=0}^{2} R_{3l}(n_l+1) \right) \right]\psi_3 + Fe^{-i\delta t} \nonumber \\
\frac{\partial}{\partial t} n_2 = -\gamma_2 n_2 +R_{32}(n_2+1)|\psi_3|^2 - R_{21}(n_1+1)n_2 - R_{20}(n_0+1)n_2 \nonumber \\
\frac{\partial}{\partial t} n_1 = -\gamma_1 n_1 +R_{31}(n_1+1)|\psi_3|^2 + R_{21}(n_1+1)n_2 - R_{10}(n_0+1)n_1 \nonumber \\
\frac{\partial}{\partial t} n_0 = -\gamma_0 n_0 +\sum_{j=1}^{3}R_{j0}(n_0+1)n_j 
\end{eqnarray}
\end{center}
where $|\psi_3|^2=n_3$. g is the polariton interaction constant, $\gamma_j$ the linewidth and $R_{jl}$ the phonon scattering rate between the state $j$ and $l$. F is the pump laser intensity. Bosonic stimulation is considered through the terms $(n_l + 1)n_j$ \cite{tassone2000}. The result of numerical simulations for the densities and a schematic of the polariton states are shown in Fig.2e and 2f respectively. The good agreement between simulations and experiments validates our interpretation for the relaxation process and the general physical principle that governs our device.\newline

Simple logic operations are achieved by exploiting the unusual emission behavior of the confined polariton states. In Fig.2b it is possible to define an intensity threshold ($I_{th}$) in order to switch on ($1$) and off ($0$) the emission from S3, S2 and S0 according to the power of the pump laser used as a control beam. The intensity threshold is set such that for $P_1$=7 mW only S3 is on. The state S2 and S0 switch on sequentially at $P_2$=9 mW and $P_3$=20 mW. Eventually at $P_4$=45 mW, S2 turns off due to bosonic stimulation. Extended logic operations can be achieved by setting $I_{th}$ such that S1 is included as an extra forth channel (Supplementary Material). The control on the polarization of the control beam allows to double the available logic operations considering the two projections of the spin up and spin down states.
Fig.3 shows the proof of principle demonstration ($a$) and a sketch of a possible implementation of the device ($c$). A laser beam tailored in intensity and polarization impinges on the polychromatic polariton laser switch. As a function of the incident beam state, different outputs can be produced and detected using a polarized beam splitter linked to sets of wavelength demultiplexers. The panels of Fig.3a are zoom of the confined polariton dispersion for different excitation powers. 
If we consider the polariton spin, nine independent configurations can be achieved by controlling the activation of the lasing from S3, S2 and S0 (see truth table in Fig.3b). The speed at which the output signal can be modulated is governed by the phonon assisted relaxation rate which is estimated to be on the order of tens of GHz. This rate will be increased in the perspective of working at higher temperature due to the enhancement of phonon-polariton scattering rate \cite{Stanley97}. 
.

\section*{Method}
\noindent {\bf Sample.} The device is engineered in a GaAs $\lambda$ semiconductor microcavity with a single 8 nm InGaAs quantum well giving a Rabi splitting of 3.5 meV. Mesas are obtained by etching the cavity before growing the top DBR. 
The reshaping of $\Delta\lambda$ of the photons wavelength (see Figure 1a) results in a confining potential of 9 meV with respect to the 2D cavity.
The detuning of the confined photon states inside the mesa with respect to the exciton is CS3 =+ 3.8 meV, CS1 = -1.7 meV and GS =- 3.7 meV, being $E_X$ = 1484.5 meV. The photonic confined state CS2, and the resulting polariton states, have been neglected from the dissertation due to their unfavorable momentum dispersion and their high excitonic content. The sample is held in a cold-finger cryostat at liquid helium temperature.\newline

\noindent {\bf Experimental Set-up.} The mesa is excited at $k_{\parallel} = 0$ with a frequency stabilized continuous-wave Ti:Sapph laser with a Gaussian profile of FWHM $\sim$ 10$\mu$m and it is prepared with circular polarization by using quarter-wave plate. The coherent emission is collected by means of a 0.5 NA microscope objective in transmission configuration. The polariton emission is then spectrally resolved and finally recorded with a high-dynamical-range CCD camera. For PLE measurements the excitation energy is varied with steps of 25 $\mu$eV. In the excitation side the power is varied by means of motorized stages.\newline

\noindent {\bf Numerical simulations.} In our model we account for the excitonic content of the states by considering a different lifetime for each of them. The values of $\gamma_j$ are then rescaled with respect to $\gamma_{0D} = \hbar/30$ ps according to the exciton content previously calculated. In the same fashion, $R_{30}$ and $R_{31}$ are scaled with respect to $R_{32}$ considering the relaxation efficiency among the states in the linear regime extracted from the PLE of Figure 1d. Experimentally these are found to be $R_{30} = R_{32}/1.7$ and $R_{31} = R_{32}/1.4$. We use $R_{32} = 5\cdot10^{-2} meV $, $R_{20} = 1.2\cdot10^{-4} meV$ and $R_{21} = R_{10} =  0.6\cdot10^{-4} meV$. Moreover this model takes into account the blueshift of S3 due to polaritons interaction inside the same state and among the others with g = 0.01 meV. This interaction between different states is introduced due to the spatial overlap between the confined states in a similar way to the interaction with the excitonic reservoir in the case of non-resonant excited polariton condensates \cite{wertz2010}.


\clearpage

\begin{figure}[h]
\includegraphics[width=1\textwidth]{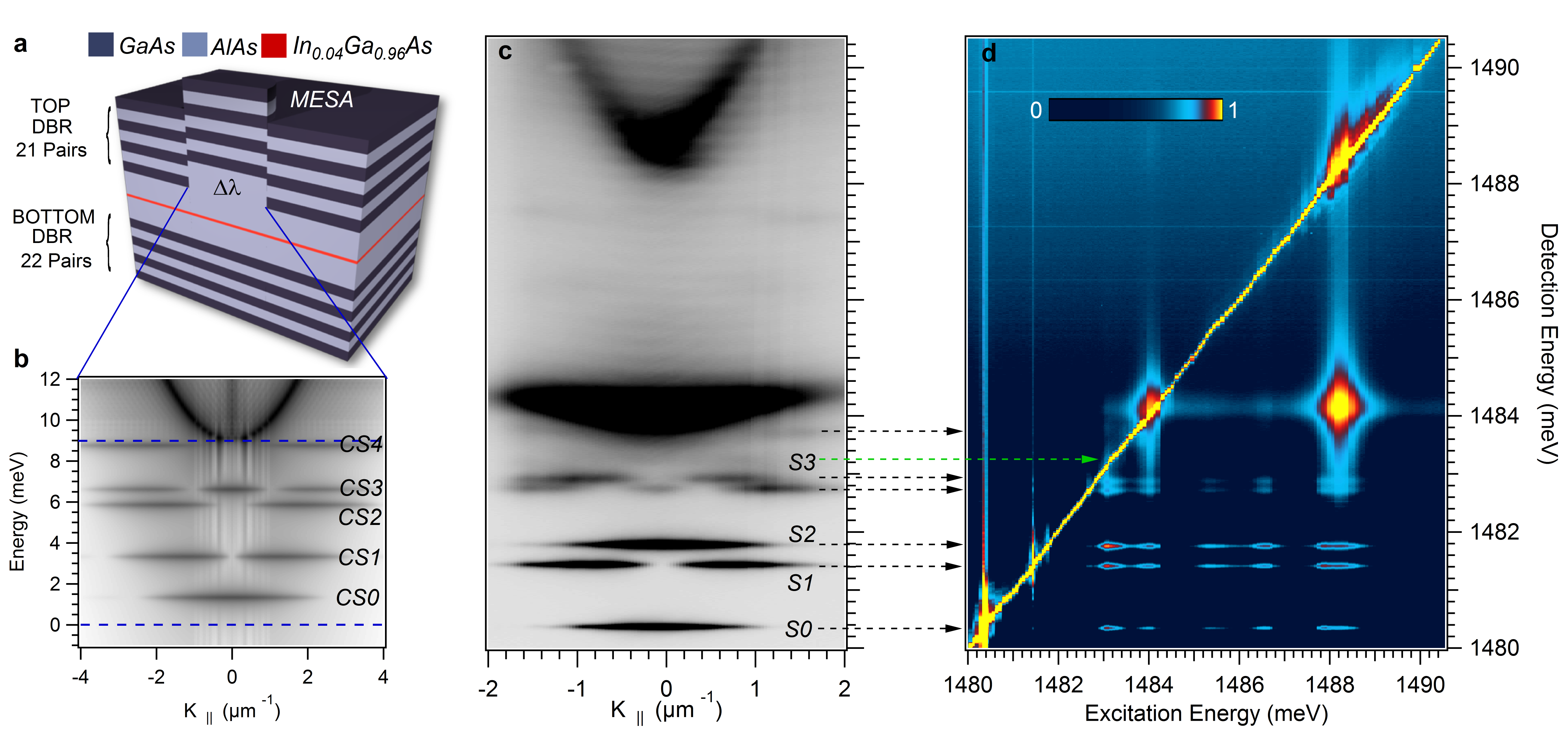}
\caption{ {\bf Sample properties in the linear regime} 
$a$ - Detailed structure of the semiconductor microcavity and the engineered photonic potential (mesa).
$b$ - Numerical simulation of the confined photonic modes (CS0-CS4) inside the mesa. The blue lines are a guide for the eyes representing the energy of the confining potential.
$c$ - Experimental dispersion for non resonant excitation. The dispersion shows the upper and lower polariton branches at strong positive detuning ($\delta_{2D}$ = 4 meV) and the lower confined states of a 3 $\mu$m diameter mesa. The states considered in the working principle of the device are label as $S_j$ with $j = 0,1,2,3$.
$d$ - Polariton emission (right axis) as a function of the excitation energy (bottom axis) for P =100 $\mu$W.
 }
\end{figure}

\clearpage
\begin{figure}[h]
\includegraphics[width=1\textwidth]{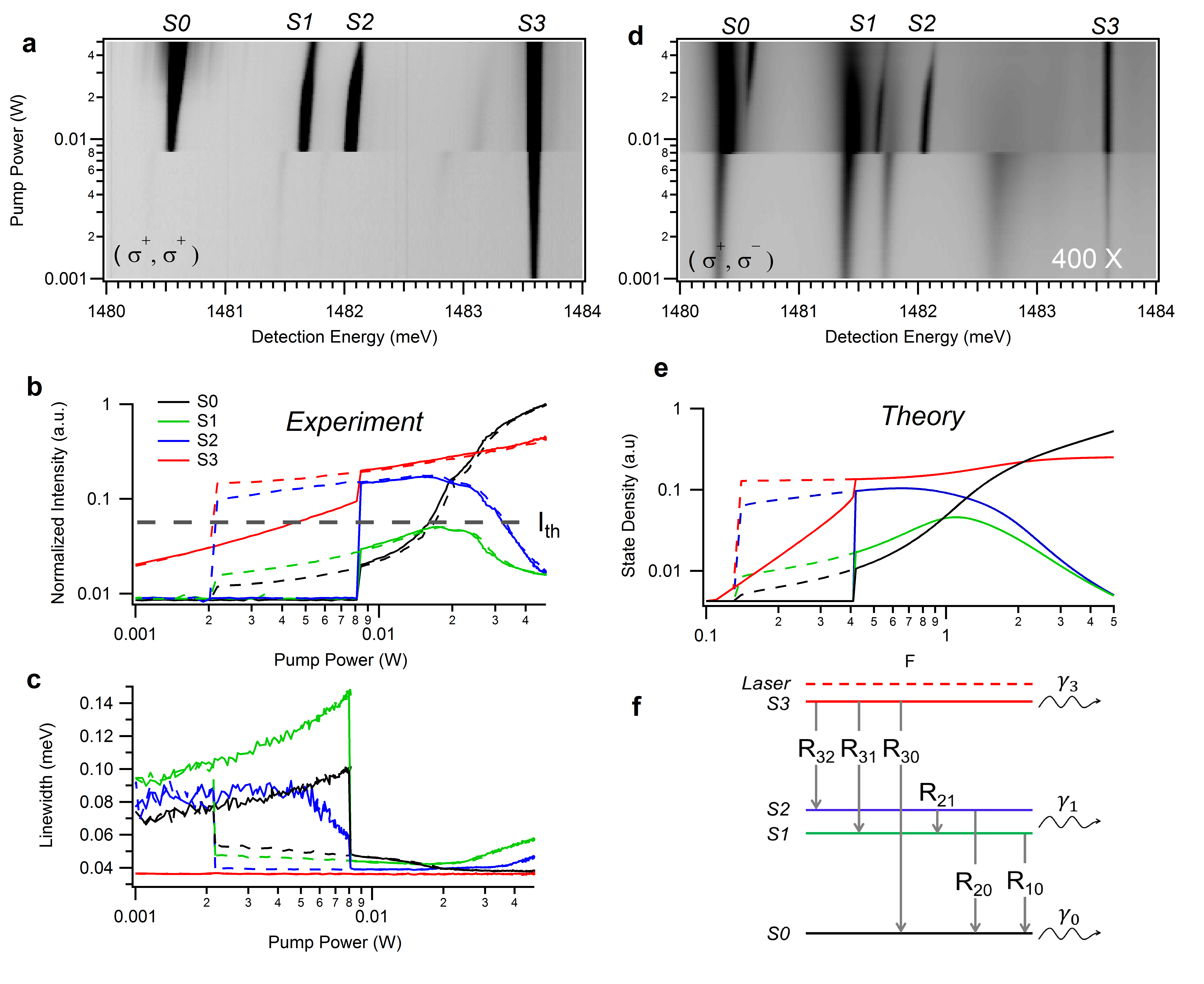}
\caption{ {\bf Nonlinear relaxation inside the mesa} 
$a$ - Spectrally resolved intensity of the mesa emission as a function of the pump power (left axis in log scale) when excited 0.3 meV above S3. The sample is excited $\sigma^+$ and the emission detected in $\sigma^+$ polarization. The color scale is saturated in order to make the low power region visible.
$b$ - Integrated and normalized intensities for all the confined polariton states as a function of the increasing (solid lines) and decreasing (dashed lines) pump power. $I_{th}$ is the intensity threshold defined for the switch operation.
$c$ - Linewidth of the polariton states as a function of the pump power.
$d$ - Spectrally resolved intensity of the mesa emission as a function of the pump power for $\sigma^+$ excitation and $\sigma^-$ detection. The color scale is 400 times more intense compared to $a$ so that the weak $\sigma^-$ emission is visible.
$e$ - Numerical simulation of the density of polariton states coupled through polariton relaxation.
$f$ - Sketch of the energy levels with the relaxation process ($R_{jl}$) and the loss ($\gamma_j$).
 }
\end{figure}

\clearpage
\begin{figure}[h]
\includegraphics[width=1\textwidth]{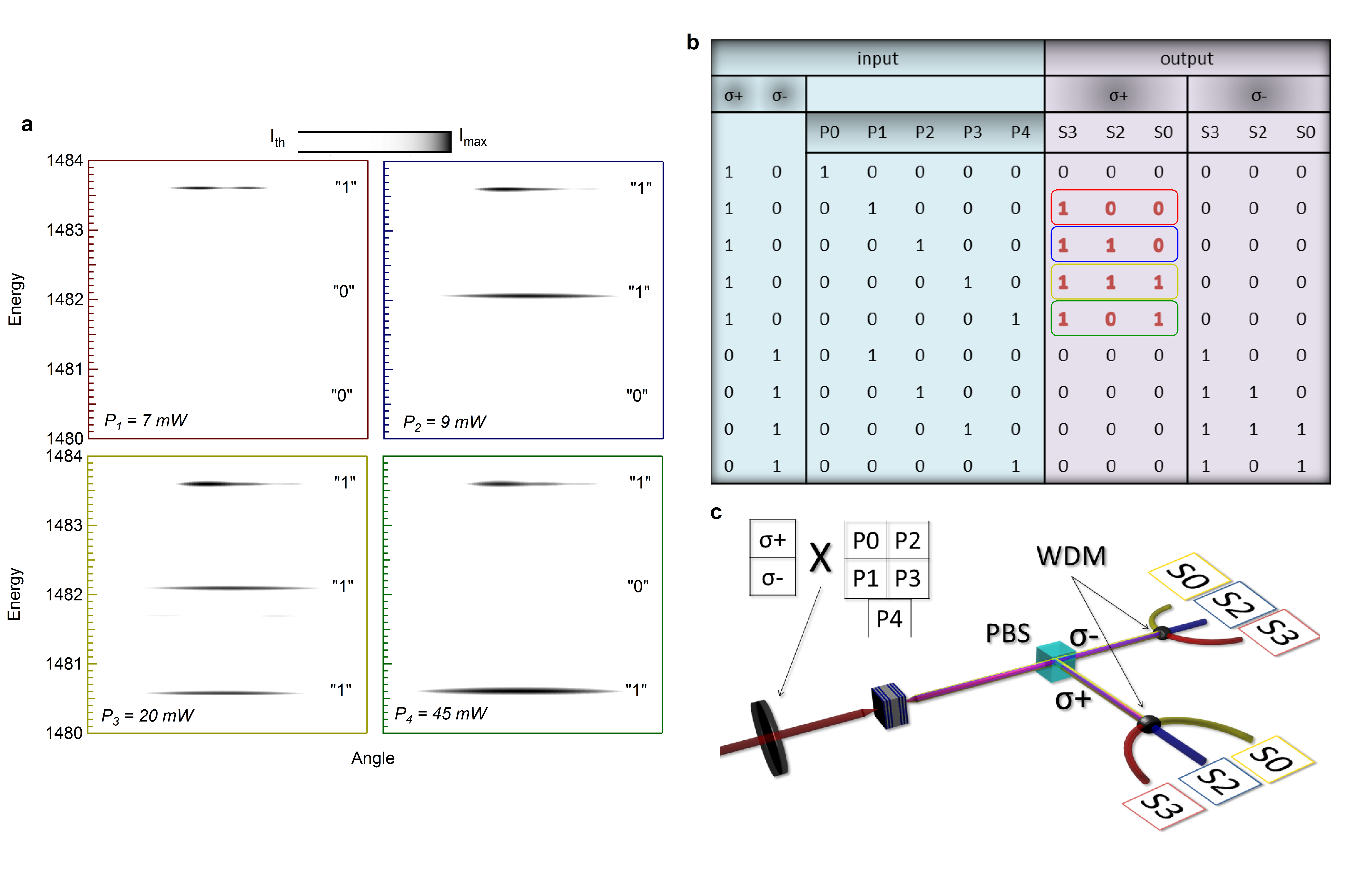}
\caption{ {\bf Proof of principle demonstration of the polariton laser switch.} 
$a$ -  The panels show the dispersion of the mesa for different powers of the control beam . The emission of three states can be controlled (0 - 1) by tailoring the input power. A color scale minimum value is set in order to simulate an intensity threshold for the switch. At $P_1$= 7mW only S3 is visible (1,0,0). By increasing the power state S2 and S0 switch on sequentially at $P_2$ = 9 mW (1,1,0) and $P_3$ = 20 mW (1,1,1). Eventually the intensity of S2 vanishes below $I_{th}$ at $P_4$ = 45 mW (1,0,1).
$b$ - Truth table representing the nine different output of the spin switch. The input parameters are the laser polarization ($\sigma^+$ and $\sigma^-$) and the power ($P_0$,$P_1$,$P_2$,$P_3$ and $P_4$). The combinations corresponding at the polariton emission of $a$ are highlighted in colors.
$c$ - Sketch of a possible implementation of the device. The polarization and the power of the control beam are tailored before the polariton injection in the sample. The multiwavelegth laser emission is then split with a Polarizer Beam Splitter (PBS) and sent to a wavelenght demultiplexer (WDM) for the reading of the six output channels.
}
\end{figure}

\clearpage

\section{Supplementary Information}

\subsection{Insight at Low power regime}
Another scan of the pump power has been performed focusing on the linear regime only,  making sure to increase the signal to noise ratio and to be able to highlight the behavior before the onset of the nonlinearity. The spectrally resolved intensity map as function of the pump power is shown in Figure 1SM. The input beam is prepared with circular $\sigma^+$ polarization and no polarization filter is applied in detection. The integrated intensity plot shows how the states emission increases linearly for all the confined states. The linewidths extracted for this regime, with $\sigma^+$ emission and detection, are plotted in Figure 1c in the main text. 

\setcounter{figure}{0} \renewcommand{\thefigure}{\arabic{figure} SM}

\begin{figure}[h]
\includegraphics[width=1\textwidth]{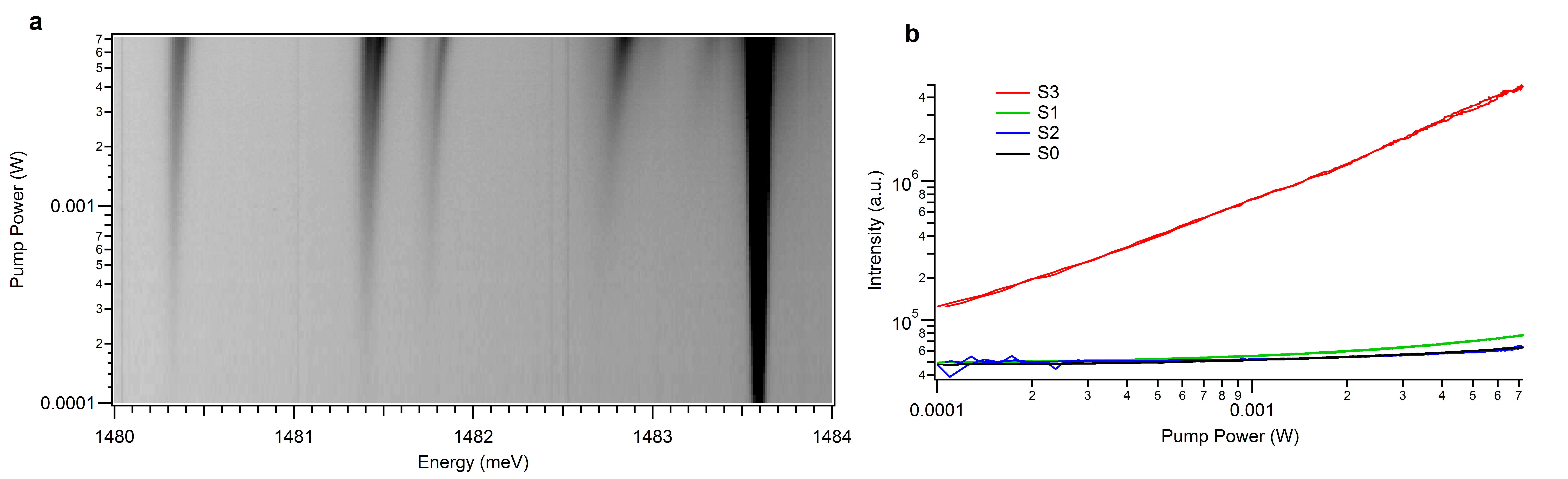}
\caption{{\bf Power scan in the low power regime} 
$a)$ Spectral resolved intensity of the mesa emission as function of the pump power (left axis in log scale) when excited 0.3 meV above $S3$. Sample is excited $\sigma^+$ and no polarization filter is applied in detection.
$b)$ Integrated intensities (left axis in log scale) for all the confined polariton states as a function of the pump power (bottom axis in log scale).
 }
\end{figure}

\clearpage

\subsection{Extended switch operations}
Extended logic operations can be achieved by decreasing the intensity threshold ($I_{th}$) for the switching.  From Figure 2SM it is clear that the state S1 can be included in the working principle adding two extra logic combinations. The proof of principle demonstration of such device is displayed in Figure 3SM. This modification produces a truth table with thirteen different outputs. Figure 4SM shows the spectral emission for the principal different operations. In this configuration, at P = 10 mW  also S1 switches ON generating a (1,1,1,0) output. At P = 20 mW the new added feature creates a (1,1,1,1) combination. 

\begin{figure}[h!]
\includegraphics[width=0.7\textwidth]{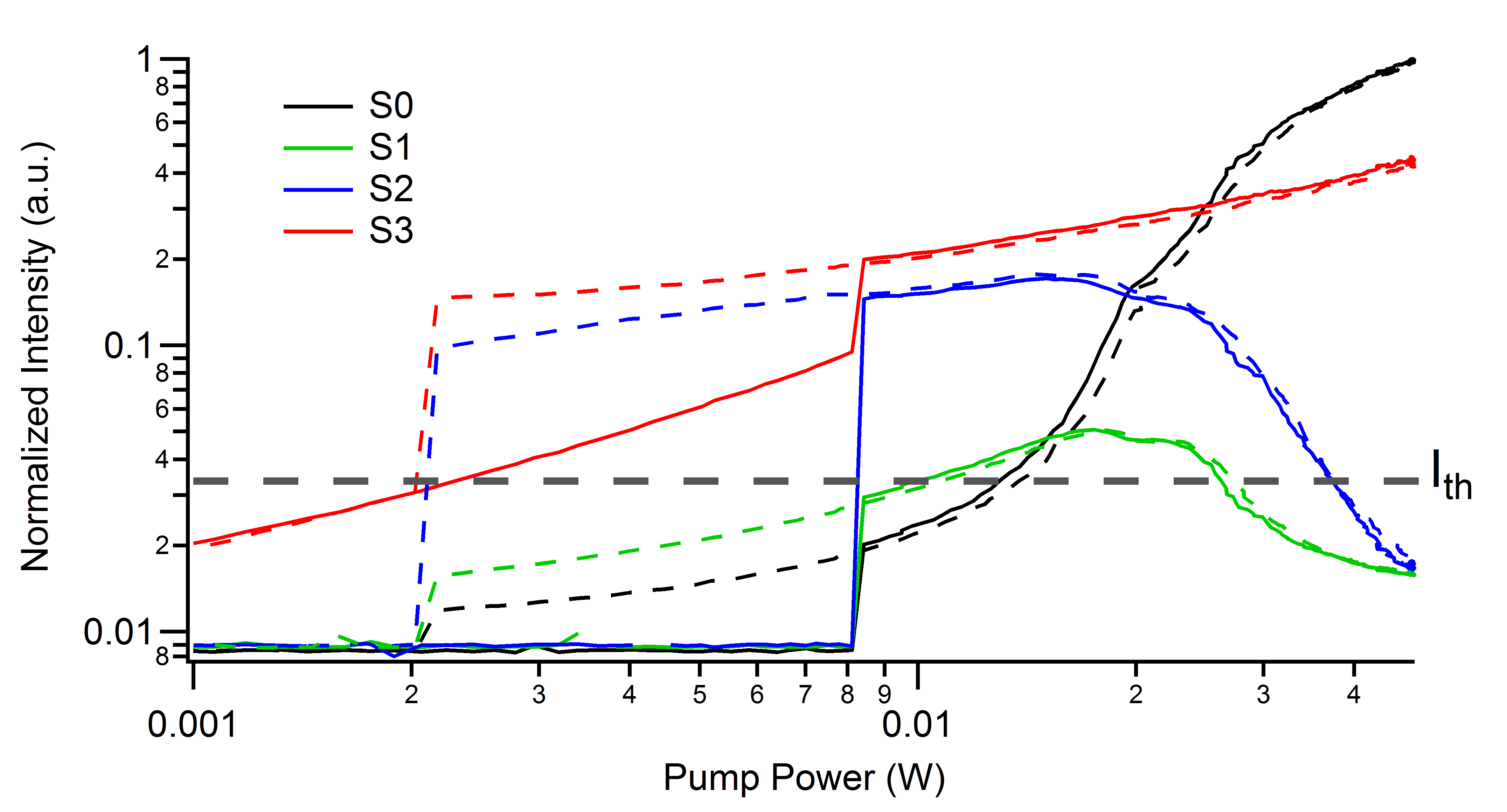}
\caption{{\bf Inclusion of S1} Integrated and normalized intensity of the confined states emission as a function of the pump power. $I_{th}$ has been lowered with respect to Figure 1b of the main text in order to include S1 in the working operation of the switch.}
\end{figure}

\begin{figure}[h!]
\includegraphics[width=0.8\textwidth]{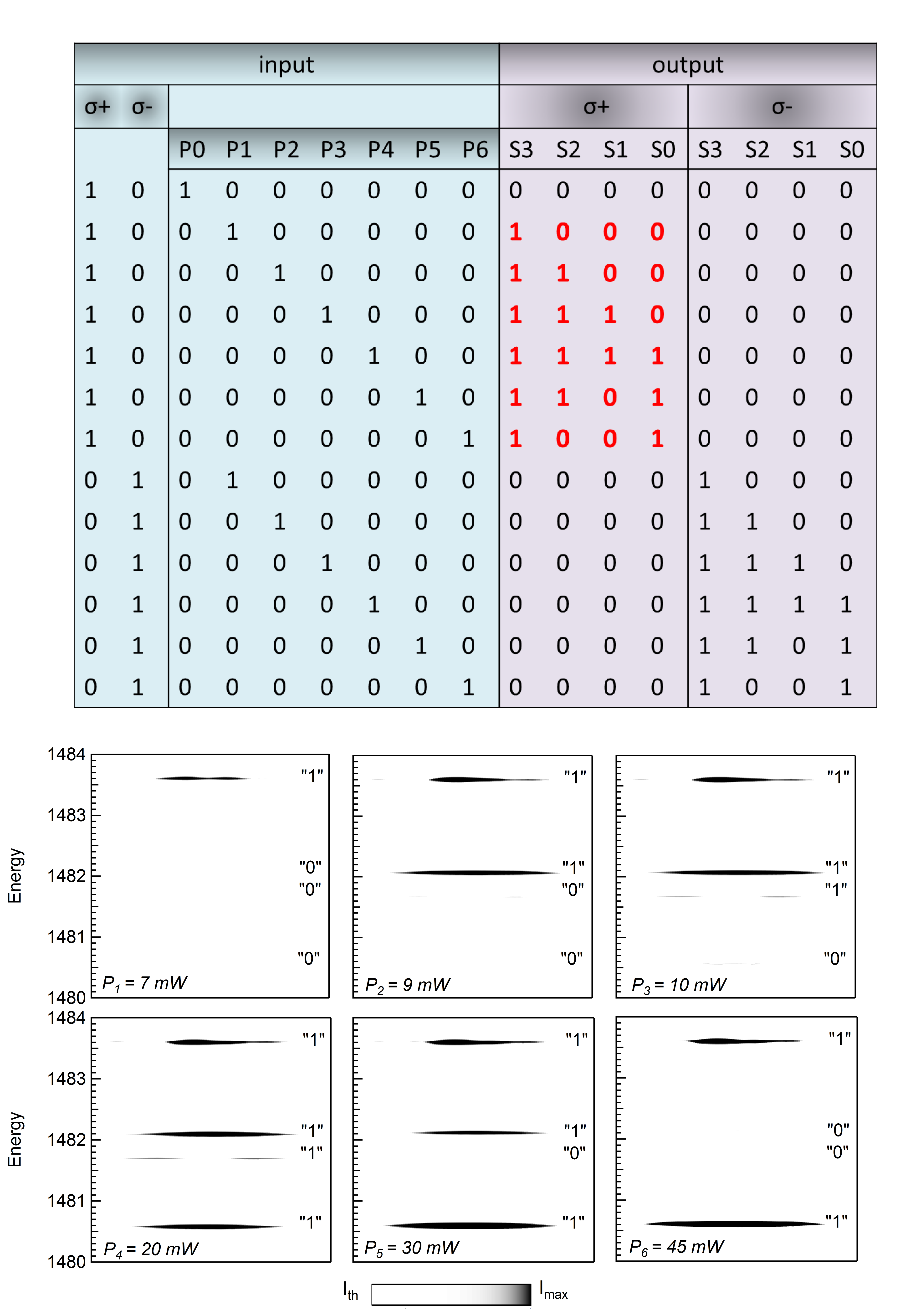}
\caption{{\bf Proof of principle demonstration of the extended polariton laser switch.} The panels show the dispersion of the mesa states for different powers of the control beam . The emission of four states can be controlled (0 / 1) by tuning the input power. Color scale minimum value is set in order to simulate an intensity threshold for the switch. At $P_1$= 7mW only S3 is visible (1,0,0,0). By increasing the power state S2, S1 and S0 switch on sequentially at $P_2$ = 9 mW (1,1,0,0), $P_3$ = 10mW (1,1,1,0) and $P_4$ = 20 mW (1,1,1,1). Eventually the intensity of S1 and S2 vanishes below $I_{th}$ at $P_5$ = 30 mW (1,1,0,1) and $P_6$ = 45 mW (1,0,0,1). The truth table for such a device is reported on the top of the figure.}
\end{figure}

\begin{figure}[b!]
\includegraphics[width=0.9\textwidth]{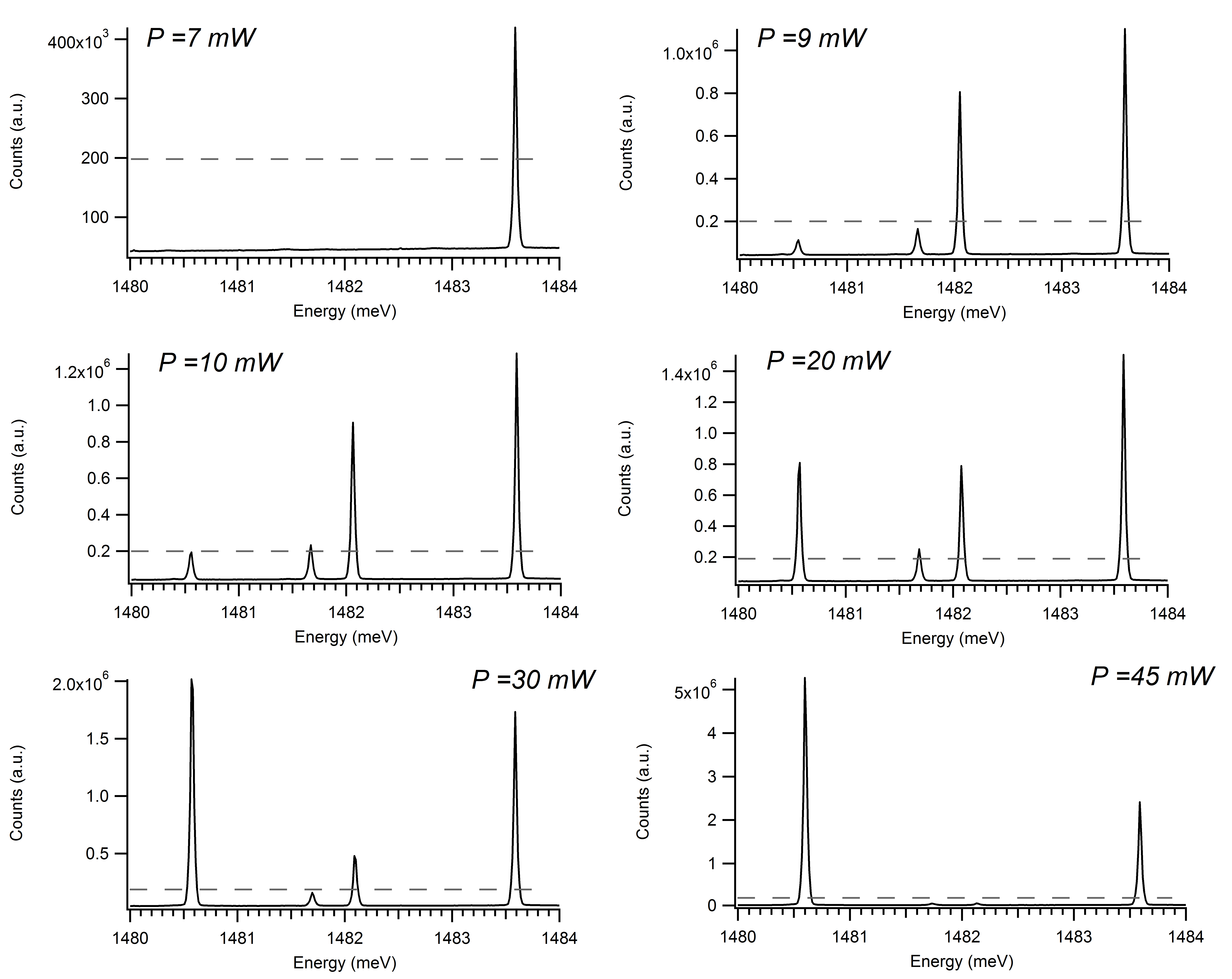}
\caption{{\bf Spectral emission for the extended polariton laser switch.} The panels show the spectra of the mesa states for different powers of the control beam. The emission of four states can be controlled (0 / 1) by controlling the input power. Dashed line corresponds at the threshold intensity for the switch.}
\end{figure}

\clearpage
\subsection{Counter-polarized detection}
We provide in this section an insight on the $\sigma^-$ emission with a $\sigma^+$ excitation. Contrarily to the co-polarized case, the linewidth of the confined states increases after the threshold. This is due to the fact the bosonic stimulation occurs only for the co-polarized component while the counter-polarized component experiences an homogeneous broadening. The broadening is the result of the enhancement of the incoherent polariton-polariton interaction while increasing the polariton population. Indeed, the spin-flip mechanism responsible for the $\sigma^-$ population is not efficient enough to reach the occupation threshold needed for the onset of the bosonic stimulation.

This phenomenon is illustrated in Figure 5SM. The information about the $\sigma^-$ component of S2 has not been extracted due to the overlap with the blueshifted $\sigma^+$ component of S1.\newline
Figure 6SM shows the spectral profiles of the confined state emissions for the main pump powers. The coexistence of the incoherent and coherent emission for each state is evident, for example, at P = 45 mW. 

\begin{figure}[h]
\includegraphics[width=0.6\textwidth]{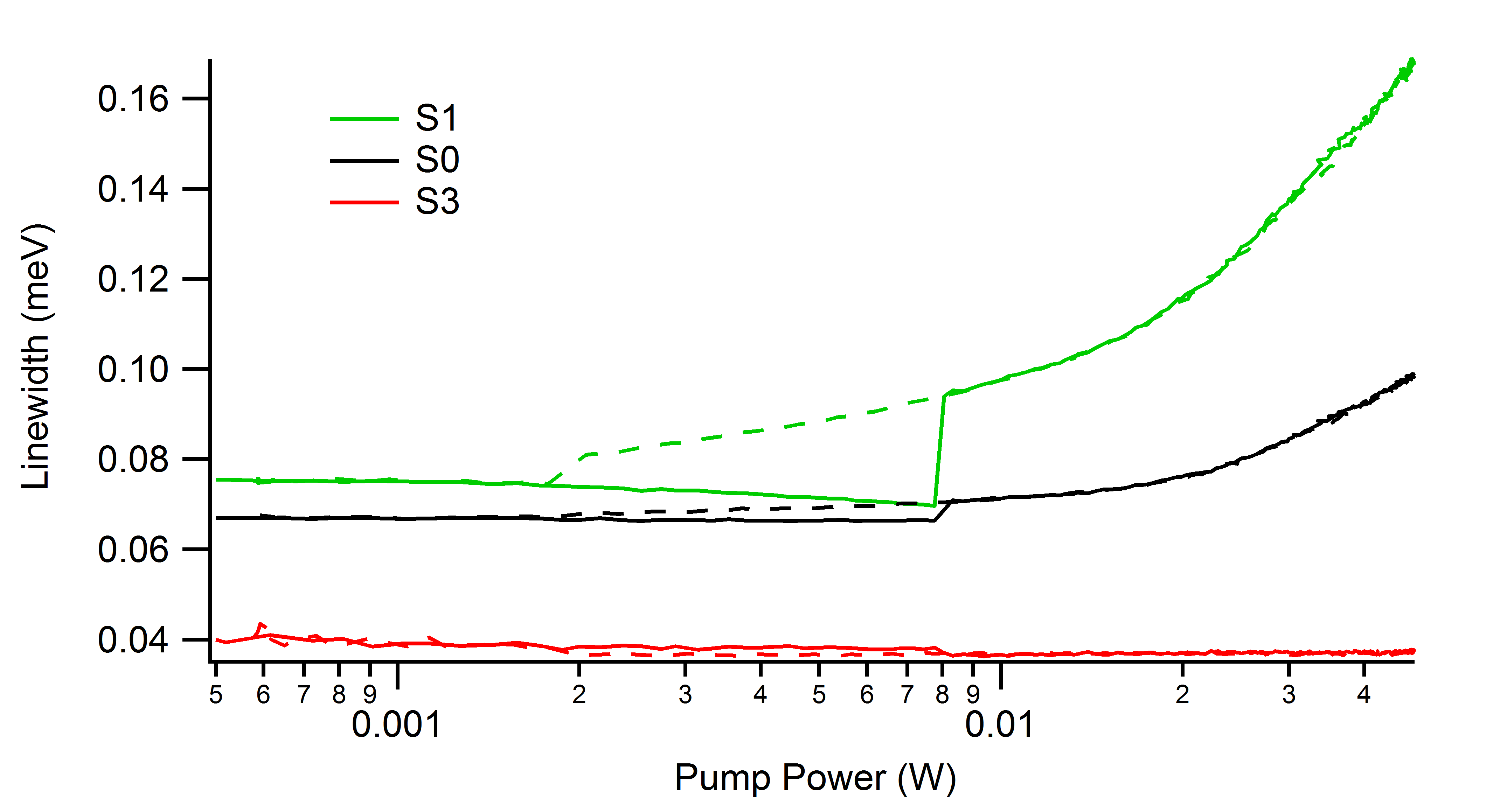}
\caption{{\bf Counter-polarized linewidth.} 
Linewidth of the confined polariton states as a function of the pump power (left axis in log scale). Sample is excited $\sigma^+$ and the emission detected in $\sigma^-$ polarization. The plots show the behavior of S3, S1 and S0. The linewidth of the confined states increases after the threshold. 
 }
\end{figure}

\begin{figure}[h]
\includegraphics[width=0.9\textwidth]{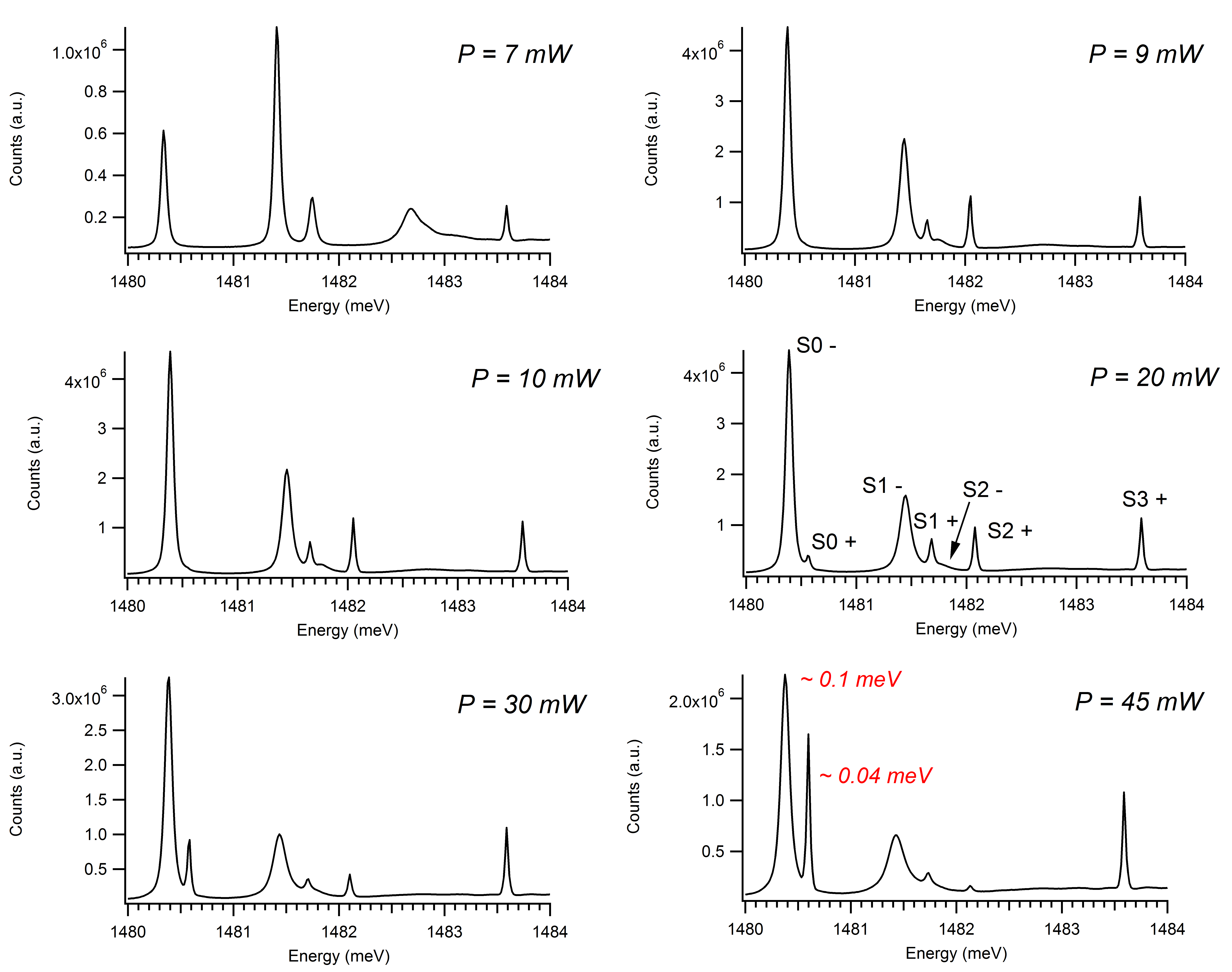}
\caption{{\bf Spectral emission in counter circular polarization.} The panels show the spectra of the mesa states for different powers of the control beam for counter-polarized circular detection. The spectra are extracted from Figure 1d in the main text and show the splitting of the two polarization components for each state. All the different states (S0-S3) with different polarization ($\sigma^+$, $\sigma^-$) are labeled in the plot at P = 20 mW. The linewidth of the $\sigma^+$ is almost half compared to $\sigma^-$ showing the coherence of the co-polarized component.}
\end{figure}

\end{document}